\documentclass[a4paper,12pt]{article}

\usepackage{amsmath,amsfonts,amssymb,epsfig}
\usepackage{slashed}
\numberwithin{equation}{section}

\usepackage{cancel}
\usepackage{cite}

\def\tr{\mathrm{tr}}

\def\beq{\begin{equation}}
\def\eeq{\end{equation}}
\def\bal{\begin{align}}
\def\eal{\end{align}}

\def\2b2[#1,#2][#3,#4]{\left( \begin{array}{cc} #1 & #2 \\ #3 & #4 \end{array}
\right)}
\def\3b3[#1,#2,#3][#4,#5,#6][#7,#8,#9]{\left( \begin{array}{ccc} #1 & #2 #3 \\
#4 & #5 & #6\\#7&#8&#9\end{array} \right)}

\newcommand{\C}[1]{\mathcal{#1}}

\def\ov{\overline}

\author{Karim~Benakli$^\clubsuit $\footnote{kbenakli@lpthe.jussieu.fr}, \, 
Mark~D.~Goodsell$^\diamondsuit $\footnote{mark.goodsell@desy.de}  \,  and  \,  Ann-Kathrin~Maier$^\spadesuit $\footnote{ann-kathrin.maier@epfl.ch} }
\date{}
\title{\vspace{-3cm}
\hfill{\small{DESY 11-061 }}\\[2cm]
Generating $\mu$ and $B\mu$ in models with Dirac Gauginos}
\begin{document}
\maketitle
\vspace{-1cm}
\begin{center}
\emph{$^\clubsuit $Laboratoire de Physique Th\'eorique et Hautes Energies,  CNRS, UPMC
Univ Paris 06
Boite 126, 4 Place Jussieu, 75252 Paris cedex 05, France \\
$^\diamondsuit$Deutsches  Elektronen-Synchrotron, DESY, Notkestra\ss e 85, 22607  Hamburg,
Germany\\
$^\spadesuit$ Laboratoire de physique de la mati\`ere complexe, SB-EPFL, CH-1015 Lausanne, Switzerland}
\end{center}
\abstract{We consider the extension of the Minimal Supersymmetric Standard Model by Dirac masses for the gauginos.
We study the possibility that the same singlet $\mathbf{S}$ that pairs up with the bino, to form a Dirac fermion, is used to generate 
$\mu$ and $B\mu$ terms through its vacuum expectation value. For this purpose, we assume that, in the Higgs potential, the necessary $R$-symmetry breaking  originates entirely from a superpotential term $\frac{\kappa}{3} \mathbf{S}^3$ and discuss the implications for the spectrum of the model.}

\newpage

\section{Introduction}

The supersymmetric extension of the standard model introduces new charged
particles that need to a acquire a large mass in order to explain the absence of
evidence in present collider experiments.In particular, the predicted  gauginos 
are fermionic states that can obtain (after supersymmetry breaking) either Majorana 
or Dirac masses. Here we are interested in this latter case \cite{Fayet:1978qc,Polchinski:1982an,Hall:1990hq,Fox:2002bu,Nelson:2002ca,Antoniadis:2005em,Antoniadis:2006eb,Antoniadis:2006uj,Hsieh:2007wq,Kribs:2007ac,Amigo:2008rc,Blechman:2009if,Benakli:
2008pg,Belanger:2009wf,Benakli:2009mk,Chun:2009zx,Benakli:2010gi,Carpenter:2010as,Chacko:2004mi,Choi:2008ub,Choi:2009jc,Choi:
2010an,Choi:2010gc,DeSimone:2010tf,Kribs:2008hq,Abel:2011dc,Davies:2011mp}.

An important feature in models with Dirac gauginos is the fate of $R$-symmetry. 
In the global supersymmetric models considered here, it appears as a continuous
symmetry (which can be broken to a discrete subgroup).  It can not be
spontaneously broken at the electroweak scale with a generic vacuum expectation
value (vev), as this would lead to a massless  
$R$-axion with a coupling insufficently suppressed to have evaded early
discovery. There remain two options: either it is conserved or explicitly
broken. In order to quantify the required size of $R$-symmetry breaking, one
needs to identify the minimal set of operators that violate the symmetry.

First, it is usual to consider that $R$-symmetry is  broken in the Minimal
Supersymmetric extension of the
Standard Model (MSSM) by Majorana gaugino masses. We can however use instead
Dirac masses for the gauginos, pairing them with additional states in adjoint
representations (henceforth DG-adjoints): a singlet $\mathbf{S}$ for $U(1)_Y$, a
triplet $\mathbf{T}$ for $SU(2)_w$ and an octet $\mathbf{O_g}$ for $SU(3)_c$.

Second, the gravitino mass required in flat space-time breaks $R$-symmetry. Again, this
can be avoided by taking a Dirac mass for the
gravitino. Such masses require the gravitational multiplet to be in  $N=2$ representations. 
To illustrate such a scenario, consider that the $N=1$ gauge and matter fields appear on 3-branes. These are
localised in a bulk having one flat extra dimension of radius $R$.  Then a
Dirac gravitino mass of size $1/2R$, and preserving $R$-symmetry,  is obtained
when the $N=2$ supergravity is  broken through a Scherk-Schwarz mechanism. 
Alternatively, the effect of $R$-symmetry being broken by minimal coupling to supergravity may be estimated and used
as a minimum estimate for Majorana masses induced in the model \cite{Abel:2011dc,Davies:2011mp}. 

Finally, the simultaneous presence of $\mu$ and $B\mu$ terms in the Higgs
sector is incompatible with $R$-symmetry. It is difficult to arrange a
viable  electroweak symmetry breaking  which preserves $R$-symmetry 
and satisfies the LEP bound. There are some interesting possibilities, such as adding extra fields \cite{Kribs:2007ac} or interactions with the supersymmetry breaking sector to generate new Higgs couplings \cite{Davies:2011mp}. However, we shall take the philosophy that, being a chiral symmetry, it is natural for $R$-symmetry to be broken in the Higgs sector. Our approach has the advantage that we do not need to introduce any interactions between the Higgs and supersymmetry-breaking sectors, and no additional mass scales.

It is popular to use the adjunction of a singlet to the MSSM as a way to address
some its issues.  One is the so-called $\mu$-problem and it is the main
motivation of the NMSSM (see \cite{Ellwanger:2009dp} and references therein).
There one starts by a vanishing tree-level $\mu$-term, as it is forbidden by a
Peccei-Quinn symmetry, and adds to the MSSM
a singlet with coupling $\mathbf{\lambda SH_u\cdot H_d}$. The potential of the
singlet is arranged such that  the breaking of supersymmetry induces a vacuum
expectation value  to $S$ naturally of the order of electroweak scale. 

Another issue is the possibility to increase the  Higgs mass. In the MSSM, the
tree level quartic Higgs coupling is given by the $SU(2)_W\times U(1)_Y$
$D$-terms. Therefore it is governed by the strength of associated gauge
couplings.  This implies that
the  lightest Higgs boson mass is bounded to be smaller than $M_Z$. Agreement
with present collider experiments is then obtained by radiative corrections. 
With
the presence of the coupling $\mathbf{\lambda SH_u\cdot H_d}$, the lightest
Higgs mass is now bounded, at tree level,  to be:
\begin{eqnarray}
m_h^2 \leq M_Z^2 (s_W^2 c^2_{2\beta} + \frac{2 \lambda^2}{(g')^2 + g^2} 
s^2_{2\beta})
\label{mhboundgenral0}
 \end{eqnarray}
where $g'$ and $g$ are the gauge couplings of the hypercharge $U(1)_Y$, and the
weak $SU(2)_W$  respectively, while $\beta$ is defined by the ratio of the two
Higgs vevs, $\tan{\beta} = \frac{<H_u>}{<H_d>}$. It is clear from
(\ref{mhboundgenral0}) that a tree-level bound larger than $M_Z$ can be obtained
for a large value of $ \lambda$, and $s^2_{2\beta} \rightarrow 1$. It can be
shown that
such a possibility is in agreement with bounds from the electroweak precision
measurements \cite{Barbieri:2006bg}.

We wish to study here if the singlet $\mathbf{S}$ that pairs up with the bino,  allowing a
Dirac soft mass, can be used in  similar ways. We will show how the addition of a cubic superpotential coupling $\frac{\kappa}{3} \mathbf{S}^3$ may indeed allow the generation of both $\mu$ and $B\mu$ terms,
 and  to push the tree-level Higgs mass above the LEP bound. It is
important to stress that although it is in many ways similar to the NMSSM, there are additional particles and couplings, and therefore a separate study is required.

The study here will be performed from an entirely low-energy perspective. However, if we consider the UV completion of the model, one might ask what symmetries allow the generation of the cubic superpotential coupling and not other similar polynomials in $\mathbf{S}$. We could suppose that the model contains different sectors obeying different symmetries, the singlet being a ``bulk'' state 
that belongs to and interacts with  all of them. These sectors are classified as

\begin{itemize}

\item a hidden or secluded sector  that breaks supersymmetry respects  $U(1)_R$ symmetry. 
Supersymmetry breaking appears as the vev of a $D$-term or  an $F$-term with zero $R$-charge. We include 
supersymmetry breaking messengers in this sector. 

\item another hidden sector which instead  respects a $\mathbb{Z}_3$ symmetry under which the singlet transforms, but
violates the $U(1)_R$ one.

\item the visible sector contains the MSSM fields, as well as the 
$SU(2)_w$ triplet and  $SU(3)_c$  octet DG-adjoints. This sector respects the $U(1)_R$ symmetry, but
violates the $\mathbb{Z}_3$ in its couplings to $\mathbf{S}$.

\end{itemize}

The interactions between the different sectors will lead to a collective breaking 
of supersymmetry, the $U(1)_R$ and the $\mathbb{Z}_3$ symmetries.

The  singlet $\mathbf{S}$ may appear either as an elementary or composite state, in which case
its coupling to visible matter can grow large rapidly with energy.  Its coupling
to the observable sector breaks the  $\mathbb{Z}_3$ symmetry by the $\mathbf{\lambda SH_u\cdot H_d}$
and Dirac gaugino couplings. On the other hand, its coupling to the  $\mathbb{Z}_3$ symmetry preserving sector 
is assumed to give rise to a superpotential  term $\frac{\kappa}{3} \mathbf{S}^3$  which represents the only
operator violating  $R$-symmetry (at tree level) relevant to the visible sector.

This note  is organized as follows. Section \ref{SEC:MODEL} introduces the notations and the
general framework for the model. The patterns of electroweak 
symmetry breaking partially discussed in
\cite{Antoniadis:2006uj,Belanger:2009wf,Choi:2010gc}, are reviewed in section \ref{Higgs}
in order to include  the effects  the coupling
$\kappa$. An important constraint on such an extension of the MSSM comes from
the $\rho$ parameter due to the vev of the $SU(2)_W$ triplet and is discussed in section \ref{SEC:TRIPLETVEV}.
The possibility of allowing a heavy Higgs is mentioned in section \ref{SEC:HEAVYHIGGS}.  In section \ref{SEC:SINGLETVEV} 
the possible generation of  $\mu$ and/or $B\mu$ terms through the vev of the
singlet $S$ is studied, and the main patterns of the resulting spectrum are
discussed, with a set of numerical examples. Conclusions are drawn in section \ref{CONCLUSIONS}.

\section{The Model}
\label{SEC:MODEL}

In order to allow Dirac masses for the gauginos, the  particle content of the  MSSM  is extended by 
states in the adjoint representations:
\begin{eqnarray}
\mathbf{S} & = & S + \sqrt{2} \theta \chi_S + \cdots  \\
\mathbf{T} & = & T  + \sqrt{2} \theta \chi_T + \cdots \\
\mathbf{O_g} & = & {O}_g  + \sqrt{2} \theta \chi_g + \cdots
\end{eqnarray}
where $S$ is a singlet, $\mathbf{O_g}$ a color octet and  $T= \sum_{a=1,2,3}
T^{(a)}$ an $SU(2)$ triplet. The latter can be written as:
 \begin{eqnarray}
T^{(1)}= T_1 \frac{\sigma^1}{2},  \qquad  T^{(2)}= T_2 \frac{\sigma^2}{2},&&   T^{(3)}= T_0 \frac{\sigma^3}{2}, \nonumber \\ T=\frac{1}{2} \begin{pmatrix}
T_0     &  \sqrt{2} T_+  \\
\sqrt{2}   T_-  &  -T_0
\end{pmatrix} ,&&  \nonumber \\ 
T_0= \frac{1}{\sqrt{2}}(T_R+i T_I),  \quad T_+=\frac{1}{\sqrt{2}}( T_{+R}+i T_{+I}), && \quad T_-=\frac{1}{\sqrt{2}}( T_{-R}+ iT_{-I}), 
\end{eqnarray}
and $\sigma^a$ are the Pauli matrices.

The  Dirac gaugino masses are described with superfields by the Lagrangian:
\begin{align}
\mathcal{L}^{Dirac}_{gaugino}=& \int  d^2\theta  \left[   \sqrt{2}
\textbf{m}^\alpha_{1D} \mathbf{W}_{1\alpha} \mathbf{S}  
+  2\sqrt{2} \textbf{m}^\alpha_{2D} \textrm{tr}(\mathbf{W}_{2 \alpha}
\mathbf{T}) +  2\sqrt{2} \textbf{m}^\alpha_{3D} \textrm{tr}(\mathbf{W}_{3
\alpha} \mathbf{O_g})  \right]  +h.c. 
\label{Newdiracgauge}
\end{align}
where we have introduced  spurion superfields
\begin{eqnarray}
\textbf{m}_{\alpha iD} &= & \theta_\alpha m_{iD}.
\end{eqnarray}
The  integration over the Grasmannian coordinates leads to
\beq
\int d^2 \theta 2\sqrt{2} m_D \theta^\alpha \tr (W_\alpha \Sigma) \supset - m_D
(\lambda_a \psi_a) + \sqrt{2} m_D \Sigma_a D_a
\eeq
Then with $D^a_b = - g_b \phi^\dagger_i R^a_b (i) \phi_i $ (where
$R_b^a (i) $ is the $a^{th}$ generator of the group $b$ in the representation of
field $i$, and $R_Y^b (i) = Y (i)$ for the hypercharge) we find
\beq
\C{L} \supset - m_{bD} \sqrt{2}  g_b \Sigma_a \phi^\dagger R^a_b \phi
\eeq

The Higgs  sector of the model is described by the superpotential:
\begin{eqnarray}
W^{(s)} &=& \mu \mathbf{H_u\!\cdot\! H_d }   + \lambda_S \mathbf{SH_u\!\cdot\!
H_d}  + 2  \lambda_T \mathbf{H_d\!\cdot\! T H_u} + \frac {M_S}{2}\mathbf{S}^2 + \frac{\kappa}{3}
\mathbf{S}^3 + M_T \textrm{tr}(\mathbf{TT}) \nonumber \\  &&+ M_O \textrm{tr}(\mathbf{O_gO_g}),
\label{NewSuperPotential}
\end{eqnarray}
the Higgs soft masses
\begin{eqnarray}
\label{potential4}
 m_{H_u}^2 |H_u|^2 +
m_{H_d}^2 |H_d|^2
 + [\tilde{B}{\mu} H_u\cdot H_d + h.c. ]
\end{eqnarray}
 as well as soft terms involving the DG-adjoint fields
\begin{eqnarray}
- \Delta\mathcal{L}^{soft} &= &  m_S^2  |S|^2 + \frac{1}{2} B_S
(S^2 + h.c.)  + 2 m_T^2 \textrm{tr}(T^\dagger T) + B_T (\textrm{tr}(T T)+ h.c.)
\nonumber \\  &&+ 
A_S  \lambda_S SH_u\cdot H_d +  2 A_T \lambda_T H_d \cdot T H_u +
\frac{1}{3} \kappa  A_{\kappa} S^3 \nonumber \\ &&+ 2 m_O^2 \textrm{tr}(O_g^\dagger O_g) 
+ B_O (\textrm{tr}(O_gO_g)+ h.c.)
\label{Lsoft-DGAdjoint}
\end{eqnarray}
with the definition $H_u\cdot H_d = H^+_uH^-_d - H^0_u H^0_d$.

In this work we will restrict for simplicity to the above terms, while the most
general renormalisable Lagrangian includes additional superpotential
interactions\footnote{Note there are no terms
$\textrm{tr}(\mathbf{T}),\textrm{tr}(\mathbf{O_g}),
\textrm{tr}(\mathbf{TTT})$ since these vanish by gauge invariance.}
\begin{align}
W^{(s)}_{2} = L \mathbf{S}  +\lambda_{ST}
\mathbf{S}\textrm{tr}(\mathbf{TT}) +\lambda_{SO} \mathbf{S}\textrm{tr}(\mathbf{O_gO_g})
 + \frac{\kappa_O}{3} \textrm{tr}(\mathbf{O_gO_gO_g}).
\label{AdjointSuperpotential}\end{align}
as well as  adjoint scalar A-terms (including the possible scalar tadpole) are given by
\begin{align}
- \Delta \mathcal{L}^{soft}_{2} =  t^S S +  \lambda_{ST} A_{ST} S \tr(TT) + 
\lambda_{SO} A_{SO} S \tr(O_gO_g) + \frac{1}{3}
\kappa_O A_{\kappa_O} \tr(O_gO_gO_g)+ h.c.   
\end{align}
and we require  $L= t^S=\lambda_{ST} = \lambda_{SO} = \kappa_O = 0$.


\section{Electroweak symmetry breaking potential}
\label{Higgs}

We define for the neutral components \cite{Belanger:2009wf,Choi:2010gc}:
\begin{eqnarray}
H^0_u  = \frac{H^0_{uR} + i H^0_{uI}}{\sqrt{2}}= \frac{1}{\sqrt{2}}[s_\beta(v+h)+c_\beta H +i (c_\beta A -s_\beta G^0)],\\ 
H^0_d  = \frac{H^0_{dR} + i H^0_{dI}}{\sqrt{2}}= \frac{1}{\sqrt{2}}[c_\beta(v+h)-s_\beta H +i (s_\beta A +c_\beta G^0)],
\end{eqnarray}
where $G^0$ is the would-be Goldstone boson (traded for the longitudinal component of the $Z$-boson). 

We shall use the compact notation:
\begin{eqnarray}
c_\beta &\equiv& \cos \beta,\qquad   s_\beta \equiv \sin \beta, \qquad  t_\beta
\equiv \tan\beta \nonumber \\
c_{2\beta} &\equiv& \cos 2\beta,\qquad   s_{2\beta} \equiv \sin 2\beta
\end{eqnarray}
Within these conventions, $v \simeq 246$ GeV, $\frac{(g')^2 + g^2}{4}
v^2 = M_Z^2$.
We are interested by the case of CP neutral vacuum, i.e. $H^0_{dI}=H^0_{uI} =0$
which implies $S_I=T_I=0$ \cite{Belanger:2009wf}. The CP-even singlet and  neutral component of the triplet may 
also acquire a vacuum expectation value, so we define
\begin{eqnarray}
S= \frac{1}{\sqrt{2}}((v_s+ S_R+i S_I) \qquad T^0= \frac{1}{\sqrt{2}}(v_T+T_R+i T_I)
\end{eqnarray}

It will also be useful to introduce the following effective mass parameters:
\begin{eqnarray}
\label{defmueff}
\tilde{\mu} & = & \mu +   \frac{1}{\sqrt{2}} (\lambda_S \,   v_S + \lambda_T  \,  v_T)
\nonumber \\
\tilde{B}{\mu} &=&B\mu +   \frac{\lambda_S}{\sqrt{2}} (M_S  + A_S) v_S+
\frac{\lambda_T}{\sqrt{2}} (M_T  + A_T) v_T +   \frac{1}{2} \lambda_S \,   \kappa  \,   v_S^2
\end{eqnarray}

\subsection{Equations of motion for the CP-even neutral fields}

The scalar potential for the CP-even neutral fields with is  given by:
\begin{eqnarray}
V_{EW} &= &\left[ \frac{g^2+g'^2}{4}c_{2\beta}^2  \,   \,   + \,   \,  \frac{
\lambda_S^2 +  \lambda_T^2}{2}s_{2\beta}^2\right] \frac{v^4}{8} \nonumber \\ 
&&+\left[  m_{H_u}^2 s_{\beta}^2+ m_{H_d}^2 c_{\beta}^2 \,   \,  +\tilde{\mu}^2
- \tilde{B}{\mu}  \,   s_{2\beta}   \,   \,  + ( g  \,    m_{2D}    \,  v_T -
g'    m_{1D}   \,   v_S   )c_{2\beta} \right] \frac{v^2}{2} \nonumber \\ 
&&+
\frac{\kappa^2}{4}v_S^4+\frac{\kappa}{\sqrt{2}}\frac{(3M_S+A_S)}{3}  \,   v_S^3+
\frac{1}{2} \tilde{m}_{SR}^2   \,   v_S^2 + \,   \,   \frac{1}{2}
\tilde{m}_{TR}^2   \,   v_T^2 
\label{VEW}
\end{eqnarray}
where the effective masses  for the real parts of the $S$ and $T$ fields read:
\begin{eqnarray}
\tilde{m}^2_{SR}& = & M_S^2+m_S^2+4 m^2_{1D}+ B_S, \qquad  \, \tilde{m}^2_{TR}= 
M_T^2+m_T^2+4 m^2_{2D}+ B_T 
 \end{eqnarray}
There is no restriction on the sign of the different  mass parameters $m_S^2$ and $B_S$ at this stage.

The imaginary parts of the fields have been dropped as their vevs are vanishing
due to the assumed CP conservation \cite{Belanger:2009wf}. The coefficients  of 
the corresponding  quadratic terms:
\begin{eqnarray}
 \tilde{m}_{SI}^2 =  M_S^2+m_S^2- B_S, \qquad  \qquad  \tilde{m}_{TI}^2=
M_T^2+m_T^2-B_T 
 \end{eqnarray}
do not, in contrast to the CP-even partners, receive contributions from
$D$-terms proportional to the Dirac masses.

As it is customarily done for the (N)MSSM, the minimization of the scalar potential allows
here also to express $\tilde{\mu}$ and $\tilde{B}{\mu} $ as a function of the other parameters:
\begin{eqnarray}
\tilde{\mu}^2+  \frac{M_{Z}^2}{2}  = \frac{ m_{H_d}^2 - t_\beta^2 \,  \, 
m_{H_u}^2}{t_\beta^2 - 1} +\left[ \frac{  t_\beta^2 +1}{t_\beta^2 -1} \right]
\left( g  \,    m_{2D}    \,  v_T - g'    m_{1D}   \,   v_S \right)
\end{eqnarray}
and
\begin{eqnarray}
M_A^2&\equiv&  \frac{2 \tilde{B}{\mu} }{s_{2\beta}}  =  2\tilde{\mu}^2+
m_{H_u}^2 + m_{H_d}^2- \frac{ \lambda_S^2 +  \lambda_T^2}{2} v^2c_{\beta}^2
\label{MA1}
\end{eqnarray}
where:
\begin{eqnarray}
v_T & \simeq &  \frac{v^2}{2\tilde{m}^2_{TR}} \ \ \left[ - g   
m_{2D}  c_{2\beta} -{\sqrt{2}} \tilde{\mu}   \lambda_T + \frac{\lambda_T
}{\sqrt{2}} ( 
M_T+    A_T)  s_{2\beta}  \right],
\end{eqnarray}
while $v_S$ is determined as a solution for the cubic equation:
\begin{eqnarray}
0 & = & \kappa^2 v_S^3+ \frac{\kappa}{\sqrt{2}}(A_\kappa +3 M_S) v_S^2-
\tilde{m}_{S0}^2 v_S + v_0^3
\end{eqnarray}
 under the assumption $\tilde{m}_{S0}^2,\tilde{m}_{S0}^2 \gg  \lambda_S \lambda_T v^2$
  \cite{Belanger:2009wf}  with 
\begin{eqnarray}
\tilde{m}_{S0}^2=- \tilde{m}_{SR}^2 - \lambda_S^2 \,  \frac{v^2}{2} + \kappa  \,
  \lambda_S \,  \frac{v^2}{2 } s_{2\beta}
 \end{eqnarray}
and 
\begin{eqnarray}
v_0^3 & = & - \frac{v^2}{2} \left[ g'    m_{1D} c_{2\beta}    -  \lambda_S  
\left( \sqrt{2} \mu - \frac{(A_\kappa +M_S)} {\sqrt{2}}s_{2\beta} +  \lambda_T
v_T  \right) \right].
\end{eqnarray}

\subsection{Masses of  the CP even neutral scalars}

The coefficient of the quadratic term  for the scalar singlet  $S_R$ is given by the effective
mass:
\begin{eqnarray}
\tilde{m}_{S}^2&=&- \tilde{m}_{S0}^2+3{\kappa^2}v_S^2+
 \frac{\sqrt{2}}{3}\kappa  \,  v_S \,   (A_\kappa +3 M_S)\\
&=&\tilde{m}_{SR}^2
+\lambda_S^2 \,  \frac{v^2}{2} - \kappa  \,   \lambda_S \,  \frac{v^2}{2 }
s_{2\beta}+3{\kappa^2}v_S^2+
 \frac{\sqrt{2}}{3}\kappa  \,  v_S \,   (A_\kappa +3 M_S)
 \end{eqnarray}
while it is
\begin{eqnarray}
\tilde{m}_T^2= \tilde{m}_{TR}^2 +\lambda_T^2  \frac{v^2}{2}
\end{eqnarray}
for the neutral component of the triplet $T^0_R$.

The resulting  CP even scalars mass matrix  takes, in the basis $\{h, H,
S_R,T^0_R\}$  the form:
\begin{eqnarray}
\left(\begin{array}{c c c c }
M_Z^2+\Delta_h s_{2\beta}^2 & \Delta_h s_{2\beta}  c_{2\beta}  & \Delta_{hs}    
& 
\Delta_{ht} \\
\Delta_h s_{2\beta}  c_{2\beta} & M_A^2-\Delta_h s_{2\beta}^2   & \Delta_{Hs}    
&\Delta_{Ht}    \\
 \Delta_{hs}     & \Delta_{Hs}      & \tilde{m}_S^2  &  \lambda_S \lambda_T
\frac{v^2}{2} \\
  \Delta_{ht}     &\Delta_{Ht}    &  \lambda_S \lambda_T \frac{v^2}{2} & 
\tilde{m}_T^2  \\
\end{array}\right) 
\end{eqnarray}

where we have introduced the compact notation:
\begin{eqnarray}
\Delta_h&=&\frac{v^2}{2}(\lambda_S^2+\lambda_T^2)-M_Z^2 
\end{eqnarray}
which vanishes when $\lambda_S$ and $\lambda_T$ take their $N=2$ values 
\cite{Antoniadis:2006uj}.
We denote non-diagonal elements describing the mixing of $S_R$ and $T^0_R$ states 
with the light Higgs $h$:
\begin{eqnarray}
 \Delta_{hs} =- 2 \frac {v_ S} {v}   \tilde{m}_{SR}^2- \sqrt{2} \kappa 
\frac{v_S^2}{v}(A_\kappa +3 M_S)- 2 \kappa^2 \frac{v_S^3}{v}, \qquad  
 \Delta_{ht} = -2 \frac {v_ T} {v}   \tilde{m}_{TR}^2
\end{eqnarray}
while
\begin{eqnarray}
 \Delta_{Hs} = g' m_{1D} v  s_{2\beta}  - \lambda_S \frac{v(A_s +M_s)}{\sqrt{2}}
c_{2\beta},\qquad
 \Delta_{Ht}  =  - g m_{2D} v  s_{2\beta}  - \lambda_T \frac{v(A_T +M_T)}{\sqrt{2}}
c_{2\beta} \nonumber \\
 \end{eqnarray}
stand for the corresponding mixing with heavier Higgs, $H$.

From this mass matrix, we see that the lightest Higgs scalar mass is bounded to be:
\begin{eqnarray}
m_h^2 \leq M_Z^2 c^2_{2\beta} + \frac{v^2}{2}(\lambda_S^2+\lambda_T^2) s^2_{2\beta}.
\label{mhboundgenral}
 \end{eqnarray}

\subsection{Masses of  the CP odd neutral scalars}

The  CP odd neutral scalars mass matrix is given, in the basis  $\{A,
S_I,T^0_I\}$,  by
\begin{eqnarray}
\left(\begin{array}{ccc}M_A^2 &  - \lambda_S v[ \frac{(M_S - A_S)}{\sqrt{2}} +k 
v_S] &-{\lambda_S v}  \frac{(A_T - M_T)}{\sqrt{2}} \\
- \lambda_S v[ \frac{(M_S - A_S)}{\sqrt{2}}  +k 
v_S] &  \tilde{m}_{aS}^2 & \lambda_S
\lambda_T \frac{v^2}{2} \\
- {\lambda_S v}  \frac{(A_T - M_T)}{\sqrt{2}}  & \lambda_S \lambda_T
\frac{v^2}{2}&  \tilde{m}_{TI}^2+  \lambda_T\frac{v^2}{2}\end{array}\right)
\end{eqnarray}

where
 \begin{eqnarray}
 \tilde{m}_{aS}^2&=& \tilde{m}_{SI}^2
+\lambda_S^2 \frac{v^2}{2}+ \kappa \lambda_S \frac{v^2}{2}
s_{2\beta}+{\kappa^2}v_s^2+
\sqrt{2}\kappa v_s (M_S- A_\kappa) 
\end{eqnarray}
is the coefficient of quadratic term in the effective Lagrangian for the
imaginary part of the singlet $S$.

\subsection{Masses of  charged scalars}

The charged would-be-Goldstone bosons, traded for the $W_\pm$ longitudinal modes,
take now the form:
\begin{eqnarray}
G^\pm_T &=&\frac{1}{\sqrt{\rho} }(G^\pm+\sqrt{2} \frac{v_T}{v}(T_\pm+(T_\mp)^*))
\end{eqnarray}
where the erstwhile Goldstone boson (before the triplet coupling is switched on) is
\begin{eqnarray}
G^+ &\equiv& c_\beta \ov{H}_d^- - s_\beta H_u^+ \nonumber\\
G^- &\equiv& c_\beta H_d^- - s_\beta \ov{H}_u^+,
\end{eqnarray}
the tree-level $\rho$-parameter given by:
\begin{equation}
\rho =1 + 4 \frac{v_t^2}{v^2}, 
\end{equation}
and the remaning orthogonal combinations of charged states can be written as
\begin{eqnarray}
T_I^{+} &=&\frac{1}{\sqrt{2} }(T_+ -(T_-)^*)\,  \,  \,  \,  
\qquad T_I^-=(T_I^+)^* \\
T_{RG}^{+}& =&\frac{1}{\sqrt{\rho} }(\frac{1}{\sqrt{2} }(T_\pm+(T_\mp)^*)-2 
\frac{v_T}{v}G^+)\qquad 
T_{RG}^{-}=(T_{RG}^+)^*.
\end{eqnarray}

The mass matrix in the Lagrangian $\mathcal{L}^{Dirac}_{gaugino}= - ({\Sigma^+})^T M_{Ch}{\Sigma^-}$ 
with $\Sigma^+ =\{ H^+, T_I^+,T_{RG}^+\}$ and $\Sigma^- =\{ H^-, T_I^-,T_{RG}^-\}$ reads then:
\begin{eqnarray}
\left(\begin{array}{ccc}\tilde{m}_{H^\pm} &  ( \frac{g^2-2\lambda_T^2}{2}v_Ts_{2\beta}- 
\frac{\lambda_T(M_T - A_T)}{\sqrt{2}})v &-\sqrt{\rho}\Delta_t 
\\
 ( \frac{g^2-2\lambda_T^2}{2}v_Ts_{2\beta}-  \frac{\lambda_T(M_T -
A_T)}{\sqrt{2}})v &  \tilde{m}_{TI}^2+\lambda_T^2 \frac{v^2}{2}+g^2v_T^2&
-\sqrt{\rho} \frac{g^2-2\lambda_T^2}{4}{v^2}c_{2\beta} \\
-\sqrt{\rho}\Delta_t  &-\sqrt{\rho} \frac{g^2-2\lambda_T^2}{4}{v^2}c_{2\beta} & 
\rho \tilde{m}_{T}^2 \end{array}\right)\nonumber \\ \nonumber \\
\end{eqnarray}
where the charged Higgs ${H^\pm}$ quadratic term reads:
\begin{eqnarray}
 \tilde{m}_{H^\pm}^2=&& M_Z^2 c_W^2 + M_A^2 + (\lambda_T^2-\lambda_S^2)\frac{v^2}{2}
+2 g m_{2D} c_{2\beta} v_T \nonumber \\&&
+2 \lambda_T^2 v_T^2
-2 \sqrt{2} \lambda_T v_T \tilde{\mu} - \sqrt{2} \lambda_T (M_T+A_T) v_T.
\end{eqnarray}

This  result is given here for completeness, but can be obtained  from \cite{Choi:2010gc} upon the use of the  new $M_A^2$, which  
contains all the dependance on $\kappa$.


\section{The $SU(2)_W$ triplet vacuum expectation value}
\label{SEC:TRIPLETVEV}

The neutral components of the DG-adjoint scalars  acquire  non-vanishing
expectation values at the minimum of the electroweak scalar potential. The minimization 
with respect to the neutral component of the triplet leads to:
\begin{eqnarray}
v_T&=&  \frac{v^2}{2(M_T^2+m_T^2+4 m^2_{2D}+ B_T)} \ \ \left[ - g   
m_{2D}  c_{2\beta} -{\sqrt{2}}  \tilde{\mu}    \lambda_T + \frac{\lambda_T
}{\sqrt{2}} ( 
M_T+    A_T)  s_{2\beta}  \right]\nonumber \\
\end{eqnarray}

As this contributes to the $W$ boson mass, the electroweak precision data give important bounds 
on the parameters of the model. For instance, using $\rho \simeq 1 +\alpha T =
1.0004^{+0.0008}_{-0.0004}$~\cite{Amsler:2008zzb},  we require:
\begin{equation}
\Delta \rho  \simeq 4 \frac{v_t^2}{v^2} \lesssim 8 \cdot 10^{-4}
\end{equation}
which is satisfied for $v_t\lesssim 3$ GeV. 

Given our assumption on having the $A$-term parameters, such as $A_T$,  small,
there are three different ways  to satisfy the bound on $v_t$:

\begin{itemize}
 \item One is to have a large supersymmetric triplet mass. In the limit
$M_T \rightarrow \infty$, the $v_T$ vanishes and the full triplet superfield  decouples. 
A Majorana mass for the winos is then required in order to avoid  the charginos being too light.

\item A second possibility is to satisfy  the bound by  taking   instead 
$m_{2D}$ large,  of the order of $\gtrsim 2$ TeV. It is also meaningful to
take simultaneously $m_T$ to be large as it is anyway expected to be in models of  gauge mediation.
This  makes not only the triplet heavy, but also the wino. In fact, in
the limit  $m_{2D}\rightarrow \infty$,  $m_T \rightarrow \infty$ and $m_T /m_{2D}
\rightarrow 0$, the weak $SU(2)$ $D$-term contribution to the effective Higgs quartic coupling cancels,
and the scalar potential becomes:
\begin{eqnarray}
V_{EW} &= &\left[ \frac{g'^2}{4}c_{2\beta}^2  \,   \,   + \,   \,  \frac{
\lambda_S^2 +  \lambda_T^2}{2}s_{2\beta}^2\right] \frac{v^4}{8} \nonumber \\ 
&&+\left[  m_{H_u}^2 s_{\beta}^2+ m_{H_d}^2 c_{\beta}^2 \,   \,  +\tilde{\mu}^2
- \tilde{B}{\mu}  \,   s_{2\beta}   \,   \,   - g'    m_{1D}   \,   v_S  
c_{2\beta} \right] \frac{v^2}{2} \nonumber \\ 
&&+
\frac{\kappa^2}{4}v_S^4+\frac{\kappa}{\sqrt{2}}\frac{(3M_S+A_S)}{3}  \,   v_S^3+
\frac{1}{2} \tilde{m}_{sR}^2   \,   v_S^2 
\label{VEW} 
\end{eqnarray}
with subsequent modification of the scalar mass matrices discussed in the previous section. 

\item Finally, the limit on $v_t$ can be satisfied just by taking  $m_T$ large
enough, keeping  $m_{2D}/m_T \rightarrow 0$. In which case, the electroweak neutral fields' 
scalar potential  becomes:
\begin{eqnarray}
V_{EW} &= &\left[ \frac{g^2+g'^2}{4}c_{2\beta}^2  \,   \,   + \,   \,  \frac{
\lambda_S^2 +  \lambda_T^2}{2}s_{2\beta}^2\right] \frac{v^4}{8} \nonumber \\ 
&&+\left[  m_{H_u}^2 s_{\beta}^2+ m_{H_d}^2 c_{\beta}^2 \,   \,  +\tilde{\mu}^2
- \tilde{B}{\mu}  \,   s_{2\beta}   \,   \,   -
g'    m_{1D}   \,   v_S   c_{2\beta} \right] \frac{v^2}{2} \nonumber \\ 
&&+
\frac{\kappa^2}{4}v_S^4+\frac{\kappa}{\sqrt{2}}\frac{(3M_S+A_S)}{3}  \,   v_S^3+
\frac{1}{2} \tilde{m}_{SR}^2   \,   v_S^2 ,
\label{VEW}
\end{eqnarray}
where we see that both contributions from the $SU(2)_W$ $D$-term and $\lambda_T$  remain.

\end{itemize}

We shall  focus in the following on the last two cases.

\section{Allowing a  heavy Higgs}
\label{SEC:HEAVYHIGGS}

From the CP-even scalar mass matrix, we see that the lightest Higgs scalar mass
is bounded by:
\begin{eqnarray}
m_h^2 \leq M_Z^2  c^2_{2\beta} + \frac{v^2}{2}(\lambda_S^2+\lambda_T^2) s^2_{2\beta} .
\label{mhboundgenral}
 \end{eqnarray}
In order to obtain a spectrum with $m_h>M_Z$, we need both a large
$\lambda_S^2+\lambda_T^2$ and
$s^2_{2\beta} \rightarrow 1$. While the second condition can be satisfied quite
easily, the 
first one requires the presence of new physics at energies $\Lambda_{NP}$ of
a few tens of TeV, and may be subject to constraints from electroweak precision
tests. In fact, as the couplings  $\lambda_T$, $\lambda_S$  grow with
energy, and now start with a large value at the electroweak scale, they
become non-pertubative in the ultraviolet very rapidly at the scale
$\Lambda_{NP}$. We expect this scale to be, at the lowest, in the range $10$--$50$ TeV 
in order to keep harmless any contributions from the new physics to
electroweak precision tests. Following the one-loop renormalization equations:
\begin{align}
16\pi^2 \frac{d}{dt} \lambda_S =& \lambda_S \bigg[ 4 \lambda_S^2 + 6 \lambda_T^2
+ 2 |\kappa_S|^2 + \cdots \bigg] \nonumber \\
16\pi^2 \frac{d}{dt} \lambda_T =& \lambda_T \bigg[ 2 \lambda_S^2 +8 \lambda_T^2
+ \cdots \bigg] \nonumber \\
16\pi^2 \frac{d}{dt} \kappa_S =&   \kappa_S \bigg[6\lambda_S^2 + 6|\kappa_S|^2
\bigg].
\end{align}
at scale $t=\log {E}$, with $E> m_{2D},m_T$,  we can consider the implications
for the three ways to obtain a heavy Higgs scenario:

\begin{enumerate}

\item  The $\mathbf{S}$-way where the coupling $\lambda_S$ is chosen to
be large enough 
while $\lambda_T\ll \lambda_S$. A bound on the size of the $\lambda_S^2 $
coupling arise from the requirement that the coupling remains perturbative up to
an energy of the order of $10$ TeV.  In such a case, the maximal value of
$\lambda_S$ at low scale ranges from approximatively $\lambda_S\sim 1.4$ for
$\Lambda_{NP}\sim 50$ TeV, to $\lambda_S\sim {2}$ for $\Lambda_{NP}\sim 10$
TeV. 

\item The $\mathbf{T}$-way, where the maximal value is somewhat lower since the
coefficient in the RGE is bigger, due to the fact that above a few TeVs all the 
triplet fields contribute. The maximal value is now smaller of order
$\lambda_{T} \sim 1$--$1.5$ for $\Lambda_{NP}\sim 10$--$50$ TeV.

\item  A combination of the two, the $\mathbf{S/T}$-way with smaller value for
each as they both run to a non-perturbative regime, with a maximal value of
$\lambda_{S,T} \sim 0.9$--$1.2$ for $\Lambda_{NP}\sim 10$--$50$ TeV.

\end{enumerate}
Note that in order to allow the largest values of $\lambda_S$, one needs to keep
$\kappa$ small as it contribute to the corresponding RGEs.

First, let us assume the presence of both $\mu$ and $B\mu$ terms not generated
by the singlet  vev $v_S$. The alternative possibility  shall be
investigated in the next section.  The  $\mu$ term by itself does not break
$R$-symmetry, but then $B_{\mu}$ does. Expecting the same source of $R$-symmetry
breaking to provide contributions to other soft terms, requires then to assume
$B_{\mu}$ and thus $M_A$ to be small.

We can consider the simplest case of a very massive singlet $\tilde{m}_S\gg v$,
so that at low energies we remain with effectively two Higgs doublets. The
resulting  CP even scalar mass matrix  becomes:
\begin{eqnarray}
\left(\begin{array}{c c }
M_Z^2 s_W^2+\Delta_h s_{2\beta}^2 & \Delta_h s_{2\beta}  c_{2\beta}  
\\
\Delta_h s_{2\beta}  c_{2\beta} & M_A^2 -\Delta_h s_{2\beta}^2      \\
\end{array}\right) 
\end{eqnarray}
which has as eigenvalues:
\begin{eqnarray}
\frac{1}{2}\left(M_A^2+M_Z^2 s_W^2 \pm \left[\Delta_h^2+(\Delta_h -M_A^2+M_Z^2
s_W^2)^2- 2 \Delta_h (\Delta_h -M_A^2+M_Z^2 s_W^2)
c_{4\beta}\right]^{1/2}\right).\nonumber \\
\end{eqnarray}

Note that the case of $\lambda$SUSY considered in \cite{Barbieri:2006bg} corresponds to neglecting the
terms $M_Z^2 s_W^2$ and leads to (for large $M_A^2$):
\begin{eqnarray}
m_{H,h} \simeq \frac{1}{2}\left(M_A^2 \pm \left[M_A^4- 2 \Delta_h(M_A^2
-\Delta_h ) c^2_{2\beta}\right]^{1/2}\right).
\end{eqnarray}

\section{The singlet vacuum expectation value}
\label{SEC:SINGLETVEV}

A scalar component  $\mathbb{S}$ with masses of a few hundred GeV does not play
an important phenomenological role, so one can assume a large $\tilde{m}_{S0}$  (as  in \cite{Barbieri:2006bg}) . 
For simplicity (and in line with our philosophy of $\kappa$ being the source of R-symmetry breaking) we will consider $M_S=0$ and negligible $A$-terms: $A_S \simeq A_\kappa \simeq 0$, which can be obtained in gauge mediated models 
with an approximate $R$-symmetric supersymmetry breaking.

For the case with $\kappa=0$ , as motivated for example some $N=2$ origin of the DG-adjoints \cite{Belanger:2009wf},
one gets:
\begin{eqnarray}
v_S &\simeq  &  \frac{v_0^3}{ \tilde{m}^2_{S0}}\nonumber \\
&\simeq &  \frac{v^2}{2(m_S^2+4 m^2_{1D}+ B_S+\lambda_S^2 \,  \frac{v^2}{2})} \ \ { \left[{\sqrt{2}} \mu   
 \lambda_S - g'    m_{1D}
c_{2\beta}  \right]}.
\end{eqnarray}

\subsection{The generation of $\mu$ and $B\mu$-terms}

We would like to switch on the trilinear coupling $\kappa$ for the singlet $S$ as the only $R$-symmetry 
breaking parameter in our model. We proceed to
investigate if a $\tilde{\mu}$ and/or $\tilde{B}{\mu} $ can be generated in a way similar to the case in the NMSSM. 
For simplicity, we take $A_\kappa$ smaller than the other relevant mass parameters.

Let us first consider the case of a large  $ \tilde{m}_{S0}$, i.e.$ \tilde{m}_{S0}^2 v_s\gg v_0^3$. 
The term $ \tilde{m}_{S0}^2 v_s$ and $\kappa^2 v_S^3$ dominate over the others in the equation
 determining the singlet. The solution can be approximated  as:
\begin{eqnarray}
 \lvert v_S \rvert &\simeq & \frac{ \tilde{m}_{S0} }{\kappa} + \frac{v_0^3}{2
\tilde{m}^2_{S0}}+\cdots \\
&\simeq &  \frac{v}{\sqrt{2}\kappa}
 (- \frac{2\tilde{m}_{SR}^2}{v^2} -\lambda_S^2 \,  + \kappa  \,   \lambda_S \, 
s_{2\beta})^{1/2} 
-\frac{ g'    m_{1D} c_{2\beta}  }{2( \frac{2\tilde{m}_{SR}^2}{v^2}
+\lambda_S^2 \,  - \kappa  \,   \lambda_S \,  s_{2\beta})}+\cdots \\
\label{vsmu}
\end{eqnarray}

 The validity of the approximation (\ref{vsmu}) requires $\tilde{m}_{S0}/\kappa$ to be larger than other masses,
   taking the effective quadratic term $\tilde{m}_{SR}^2$ to be governed by 
a sufficiently large $\lvert B_S \rvert$, 
\begin{eqnarray}
- B_S = \lvert B_S \rvert \gg m_{1D}^2
\end{eqnarray}
This hierarchy is not unnatural, it is even  quite generic in models of gauge mediation with Dirac gauginos \cite{Benakli:2010gi}. It is easy to find a generic set  of messengers with couplings to the singlet that  leads to large contribution to $\lvert B_S \rvert$ with the desired sign, following the results of  \cite{Benakli:2008pg}. Therefore:
\begin{eqnarray}
\lvert v_S \rvert &\simeq   \frac{\lvert \tilde{m}_{SR}\rvert}{\kappa} .
\label{vsmuLM}
\end{eqnarray}

 This singlet vev (\ref{vsmu}) induces both a $\mu$-term 
\begin{eqnarray}
\tilde{\mu}^2  = \frac{  \lambda_S ^2}{2} v_S^2 \simeq  \frac{  \lambda_S^2}{2
\kappa^2}(- \frac{2\tilde{m}_{SR}^2}{v^2} ) v^2
 \simeq   \frac{  \lambda_S^2}{\kappa^2}(\lvert B_S \rvert -m_S^2- 4 m^2_{1D})
\label{Approxeffectivemu}\end{eqnarray}
and a $B\mu$-term
\begin{eqnarray}
 \tilde{B}{\mu} &\simeq&\frac{\kappa}{\lambda_S}  \tilde{\mu}^2 
\simeq   \frac{  \lambda_S}{\kappa}(\lvert B_S \rvert -m_S^2- 4 m^2_{1D})
\end{eqnarray}
with sizes controlled by $\lambda_S$, $\kappa$ and $(\lvert B_S \rvert -m_S^2)$. With four parameters we can obviously fit the desired 
values for $\tilde{\mu}$. When $\kappa \lesssim \lambda_S$, this  implies that   $B\mu \lesssim {\mu}^2$. A small $R$-symmetry breaking corresponds then to a small $B\mu$, and  a hierarchy between  $B\mu$ and ${\mu}^2$ would require some amount of  tuning between the parameters in  (\ref{MA1}).

Before discussing the implications for the scalar spectrum, it is important to notice that the term $\frac {\kappa}{3} \mathbf{S}^3$ induces through the vev of $S$ a Majorana mass for the singlino $\chi_S$. 
The neutralino mass matrix, in the basis $\tilde{S}, \tilde{B}, \tilde{T}^0, \tilde{W}^0, \tilde{H}_d^0, \tilde{H}_u^0$ reads:
 \begin{equation}
{\scriptsize \left(\begin{array}{c c c c c c}
M_S + \sqrt{2} \kappa v_S  & m_{1D} & 0     & 0     & - \frac{ \sqrt{2} \lambda_S }{g_Y}M_Z s_W s_\beta &  -  \frac{ \sqrt{2} \lambda_S }{g_Y}M_Z s_W c_\beta  \\
m_{1D} & M_1   & 0     & 0     & -M_Z s_W c_\beta &   M_Z s_W s_\beta  \\
0     & 0     & M_T  & m_{2D} & - \frac{ \sqrt{2} \lambda_T  }{g_2}M_Z c_W s_\beta & - \frac{ \sqrt{2} \lambda_T  }{g_2}M_Z c_W c_\beta  \\
0     & 0     & m_{2D} & M_2   &  M_Z c_W c_\beta & - M_Z c_W s_\beta  \\
-\frac{ \sqrt{2} \lambda_S }{g_Y}M_Z s_W s_\beta & -M_Z s_W c_\beta & -\frac{ \sqrt{2} \lambda_T  }{g_2}M_Z c_W s_\beta &  M_Z c_W c_\beta & 0    & -\tilde{\mu} \\
-\frac{ \sqrt{2} \lambda_S }{g'}M_Z s_W c_\beta &  M_Z s_W s_\beta & -\frac{ \sqrt{2} \lambda_T  }{g_2}M_Z c_W c_\beta & -M_Z c_W s_\beta & -\tilde{\mu} & 0    \\
\end{array}\right) }
\label{diracgauginos_NeutralinoMassarray}
\end{equation}
Therefore a Majorana mass of order:
\begin{eqnarray}
M'_1= \sqrt{2} \kappa v_S \simeq  \frac{  2 \kappa}{\lambda_S}\tilde{\mu}
\end{eqnarray}
 spoils the  pseudo-Dirac nature of the bino, unless we have 
\begin{eqnarray}
  \frac{ \kappa}{\lambda_S} \frac{\tilde{\mu}}{m_{1D}}\ll 1.
\end{eqnarray}
Noting (\ref{Approxeffectivemu}) we see that this requires
\begin{align}
\left(\frac{\lvert B_S \rvert -m_S^2}{m_{1D}^2}\right) - 4 \ll 1,
\end{align}
which represents a degree of fine-tuning. Moreover, it requires a hierarchy between the two couplings $\kappa\ll \lambda_S$ to avoid the neutralino being mostly Higgsino (and rendering the pseudo-Dirac mass for the Bino irrelevant) which can in turn cause the pseudoscalar Higgs to be light. Hence the most generic situation would be that the neutralino contains a Majorana mix of the Bino and singlino.

The chargino masses, $- \frac{1}{2} ( (v^-)^T M_{Ch} v^+ + (v^+)^T M_{Ch}^T v^- + h.c)$
in the basis $v^+ = (\tilde{T}^+,\tilde{W}^+,\tilde{H}^+_u)$, 
$v^- = (\tilde{T}^-,\tilde{W}^-,\tilde{H}^-_d)$,
are given by
\begin{equation}
M_{Ch} = 
\left(\begin{array}{c c c}
M_T   & m_{2D} + gv_T &\frac{ {2} \lambda_T  }{g} M_Z c_W c_\beta \\
m_{2D} - gv_T & M_2   & \sqrt{2} M_Z c_W s_\beta \\
- \frac{ {2} \lambda_T  }{g} M_Z c_W s_\beta & \sqrt{2} M_Z c_W c_\beta & \tilde{\mu} - \sqrt{2}\lambda_T v_T \\
\end{array}\right) 
\label{diracgauginos_CharginoMassarray}
\end{equation}
and do not depend on $\kappa$.


\subsection{The  scalar spectrum}

The quadratic term coefficient for CP even scalar singlet $S_R$  is given by the effective
mass:
\begin{eqnarray}
{ \tilde{m}^2_{S} }\simeq 2{ \tilde{m}^2_{S0} } \simeq - 2\tilde{m}^2_{SR}\simeq 2 (\lvert B_S \rvert -m_S^2- 4 m^2_{1D})>0
 \end{eqnarray}
while the one for CP-odd $S_I$  reads:
\begin{eqnarray}
 \tilde{m}_{aS}^2
&=&  2 \lvert B_S \rvert -4m_{1D}^2 + 2 \kappa \lambda_S \frac{v^2}{2} s_{2\beta}.
\end{eqnarray}
On the other hand the coefficients of the quadratic terms of $T_R$ and $T_I^0$ are given by: 
\begin{eqnarray}
\tilde{m}_T^2 \simeq m_T^2+4 m^2_{2D}+ B_T
\end{eqnarray}
\begin{eqnarray}
 \tilde{m}_{aT}^2  \simeq m_T^2-B_T .
 \end{eqnarray}

The  CP odd neutral scalar mass  is given by
\begin{eqnarray}
 M_A^2 = \frac{2\tilde{B}{\mu} }{s_{2\beta}}&\simeq&  \frac{ 2 \lambda_S}{s_{2\beta} \kappa}(\lvert B_S \rvert -m_S^2- 4 m^2_{1D}).
 \end{eqnarray}

The charged Higgs mass is
\begin{eqnarray}
 \tilde{m}_{H^\pm}^2 &=& M_W^2 {s_{2\beta}}^2 + M_A^2 + (\lambda_T^2-\lambda_S^2)\frac{v^2}{2}.
\end{eqnarray}

The off diagonal elements describing the mixing of $S_R$, $T_R^0$ components
with the Higgs scalars $h,H$ can be approximated as:
\begin{eqnarray}
 \Delta_{hs} \simeq - \sqrt{2}(\lambda_S + \kappa s_{2\beta} )  \tilde{\mu} v ,
\qquad  
 \Delta_{ht} \simeq \sqrt{2}\lambda_T  \tilde{\mu} v+g m_{2D} v  c_{2\beta}
\end{eqnarray}
\begin{eqnarray}
 \Delta_s \simeq g' m_{1D} v  s_{2\beta},\qquad
 \Delta_t  \simeq  - g m_{2D} v  s_{2\beta} \end{eqnarray}
which means that neglecting mixings requires working  with approximations of  
order $\mathcal{O} (v \tilde{\mu} /\tilde{m}_S^2) $ for $S_R$ and
$\mathcal{O}(v m_{2D} /\tilde{m}_T^2) $ for $T_R^0$ .

With both $\mu$ and $B_\mu$ generated by the singlet vev, we can look for illustrative examples, and we would in particular like to exhibit the possibility of a larger tree-level mass for the lightest Higgs. A set of such examples is given in table \ref{model1}. The first two models have a large Higgs mass  thanks to a larger value of $\lambda_S$. To avoid a light pseudo-scalar Higgs we  required $\kappa \sim \lambda_S$. The list of gauge and Higgs sector soft masses can be read in  the table, notably with a lightest CP-even neutral Higgs mass evading the LEP bound already at tree-level. For model I, $\lambda_S = \kappa = 1.2, \lambda_T =0.1, \tan \beta = 1$ lead to the lightest Higgs mass of  $202$ GeV at tree level, while $\lambda_S =0.8,  \kappa = 06, \lambda_T =0.1, \tan \beta = 1.38$ lead to a tree-level mass of 116 GeV. We note that the spectrum is  sensitive to the relative values of the couplings $\lambda_S ,  \kappa, \lambda_T , \tan \beta $. In model III, we give an example with large $\lambda_T$ and small $\lambda_S$.

Among noticeable features that can be seen  in the presented spectra is that, due to the choice of a large $M_A$, we have $M_{H^\pm}\sim M_H\sim M_A$. We have also chosen in these examples to suppress the triplet expectation value by a large soft mass for the scalar triplet field. In 
model I, the lightest neutralino is made of  approximatively 72 percent of bino, 18 percent of singlino, while they become  approximatively 75 percent bino, 21 percent singlino in model II and nearly all the rest Higgsino. Instead, in model III, the lightest neutralino  is made of Higgsinos for about 87 percent, 7 percent wino and 4 percent  bino.

\begin{table}[!h]\begin{center}
\begin{tabular}{|c|c|c|c|} \hline
Input parameter & Model I& Model II & Model III \\\hline
$\lambda_S$ & $1.2$ & $0.8$& $0.1$\\ 
$\lambda_T $ & $0.1$ & $0.1$ & $0.7$\\ 
$\kappa$ & $1.2$& $0.6$ & $0.2$\\ 
$\tan \beta$ & $1$ & $1.38$ & $1.38$ \\ 
$m_S^2$ & $10^5$ GeV$^2$& $10^5$ GeV$^2$& $10^5$ GeV$^2$\\ 
$B_S$ &$ -10^6$ GeV$^2$& $-10^6$ GeV$^2$& $-10^6$ GeV$^2$\\ 
$m_T^2$ & 4 $10^6 $ GeV$^2$& 4 $10^6$  GeV$^2$& 4 $10^6$  GeV$^2$\\ 
$B_T$ & 0 & 0& 0\\
$A_\kappa$ & 0 & 0 & 0\\
$ m_{H_u}$ & 197 GeV & 479 GeV & 596 i GeV\\
$m_{H_d}$ & 287 i GeV& 339 i GeV & 642  GeV\\
$m_{1D}$ & 400 GeV & 400 GeV & 400 GeV\\ 
$m_{2D}$ & 600 GeV& 600 GeV& 800 GeV\\\hline
Output parameter & Model I& Model II& Model III\\ \hline
$v_S$ & -425 GeV & 838 GeV& 2548 GeV\\
$v_T$ & 0.3 GeV& 0.3 GeV& -0,08 GeV \\
$\Delta\rho$ &  5.3 $\times 10^{-6}$ &  5.9 $\times 10^{-6}$&  4 $\times 10^{-7}$\\
$\tilde{\mu}$ & -361 GeV& 474 GeV & 180 GeV\\\hline
\end{tabular}
\begin{tabular}{|c|c|c|} \hline
&Chargino masses: &612, 604, 352 GeV\\
&Neutralino masses: &740, 613, 606, 388, 352, 203 GeV\\
Model I\,  &Neutral scalars: &2332, 723, 467, 208 GeV\\
&Neutral pseudoscalars: &2001, 1211, 491 GeV\\
&Charged scalars: &2333, 2000, 471 GeV \\\hline
\end{tabular}
\begin{tabular}{|c|c|c|} \hline
&Chargino masses: &622, 602, 455 GeV\\
&Neutralino masses: &732, 619, 605, 484, 456, 215 GeV\\
Model II&Neutral scalars: &2333, 718, 580, 116 GeV\\
&Neutral pseudoscalars: &2001, 1181, 588 GeV\\
&Charged scalars: &2333, 2000, 583 GeV \\\hline
\end{tabular}
\begin{tabular}{|c|c|c|} \hline
&Chargino masses: &812, 808, 178 GeV\\
&Neutralino masses: &842, 830, 730, 226, 189, 171 GeV\\
Model III&Neutral scalars: &2564, 722, 354, 120 GeV\\
&Neutral pseudoscalars: &2005, 1166, 369 GeV\\
&Charged scalars: &2565, 2004, 394 GeV \\\hline
\end{tabular}
\end{center}

\caption{Examples of model parameters.}
\label{model1}\end{table}

An alternative approach, more easily compatible with models perturbative up to the GUT scale such as those of \cite{Benakli:2010gi}, is to take the couplings $\lambda_S \sim \kappa, \lambda_T$ to be small so that $\lambda_S, \lambda_T$ will not drastically contribute to the running of the Higgs masses. However in this case we can no longer rely on the tree level potential,  as after electroweak symmetry breaking the Higgs mass will not exceed the LEP bounds. Instead we must include, as usual, loop corrections to the Higgs potential and work at relatively large $\tan \beta$. We give an example of this in table \ref{model2}, taking squarks at $1100$ GeV. In this case, we have an almost entirely bino--singlino neutralino, with composition $81\%$--$18\%$. 

\begin{table}[!h]\begin{center}
\begin{tabular}{|c|c|} \hline
Input parameter & Model IV \\\hline
$\lambda_S$ & 0.13 \\
$\lambda_T$ & 0.13 \\
$\kappa$ & 0.13 \\
$\tan \beta$ & 31.8 \\
$m_S^2$ & 400000 GeV$^2$\\
$B_S$ & -1500000 GeV$^2$\\
$m_T^2$ & 6000000 GeV$^2$ \\
$B_T$ & 0 \\
$A_\kappa$ & 0 \\
$m_{H_u}$ & 718 GeV \\
$m_{H_d}$ & 2520 GeV \\
$m_{1D}$ & 400 GeV\\
$m_{2D}$ & 500 GeV \\
Output parameter & Model IV \\ \hline
$v_S$ & -5220 GeV \\
$v_T$ & 1.78 GeV \\
$\Delta\rho$ & 0.00021 \\
$\tilde{\mu}$ & -480 GeV \\\hline
\end{tabular}
\begin{tabular}{|c|c|} \hline
Chargino masses: &559.828, 500.624, 423.868 GeV\\
Neutralino masses: &864.145, 553.39, 552.476, 430.067, 428.46, 184.964 GeV\\
Neutral scalars: &2707.7, 2646.11, 960.048, 118.035 GeV\\
Neutral pseudoscalars: &2707.68, 2449.59, 1536.49 GeV\\
Charged scalars: &2708.7, 2646.12, 2449.59 GeV\\\hline
\end{tabular}
\end{center}
\caption{An example with small $\lambda_S, \lambda_T, \kappa$, including top-stop loop corrections.}
\label{model2}\end{table}

\section{Conclusions}
\label{CONCLUSIONS}

In any extension of the MSSM, the presence of a singlet $S$ immediately raises a question about the possibility
to use it to generate $\mu$ and $B\mu$  parameters. The $S$ vev, easily computed,  
was found in all previous studies to be typically too small.  As the combination of   $\mu$ and $B\mu$ break the $U(1)_R$  $R$-symmetry, their generation is related to the way this breaking is implemented in the potential of $S$. We have argued for
the adjunction to the model of a superpotential term $\frac{1}{3}\kappa \mathbf{S}^3$, and shown that it is sufficient. Both $\mu$ and $B\mu$ 
are induced with a ratio  $B\mu/ \mu \sim \kappa/\lambda_S$, which should be close to unity in order to obtain a successful 
electroweak symmetry breaking without excessive tuning of the parameters. We have also verified that this allows to raise the tree-level mass for the lightest CP-even Higgs above the LEP bound. This can be achieved by moderate values of either of $\lambda_S$, $\lambda_T$ as given in the explicit examples.

While this work exhibited the main features of the model, some important  issues need to be investigated before deriving phenomenological implications of it for the LHC and dark matter search experiments. Foremost among these is a thorough study of the impact of the presence of the singlet and the triplet on electroweak precision tests, such as calculating the $S$ and $T$ parameters. We believe that this merits an independent careful study by itself in the light of the disagreements in the literature on the correct treatment and the result of such an analysis already at the level of  the  (non-supersymmetric) extension of the standard model with  triplets   (see for example \cite{Chankowski:2006hs,Chen:2008jg,Skiba:2010xn}). Such an analysis will allow the determination of the regions allowed for the parameters $\lambda_S$, $\lambda_T$, $m_{2D}$ and $m_T$ beyond the tree level considerations discussed here. On the other hand, top-down models for hidden sectors that allow the generation of the $\kappa \mathbf{S}^3$, and investigation of the resulting pattern of the soft-terms are needed. We will return to these important issues in future work.

\section*{Acknowledgments}

This work is supported in part by the European contract ``UNILHC''
PITN-GA-2009-237920. MDG is supported by the German Science Foundation (DFG)
under SFB 676. AKM would like to thank the Laboratoire de Physique Th\'eorique des hautes Energies (LPTHE), Paris 
for financial support and hospitality during her masters degree when she worked on this project, MDG would like to thank
 the Centre de Physique Th\'eorique de l'Ecole Polytechnique, where part of this work was completed, for
hospitality; KB thanks Roberto Franceschini and Pietro Slavich for discussions.

\end{document}